\documentclass[12pt,onecolumn,prc,showpacs,preprintnumber,amsmath,
floatfix,amssymb]{revtex4}
\usepackage{epsfig}
\usepackage{graphicx}
\usepackage{dcolumn}
\usepackage{bm}
\def\r{\mbox{{\bf  r}}}
\def\p{\mbox{\boldmath $p$}}
\def\q{\mbox{\boldmath $q$}}
\def\k{\mbox{\boldmath $k$}}
\def\t{\mbox{\boldmath $t$}}
\allowdisplaybreaks[4]

\begin{document}
\title{Quasi-elastic neutrino charged-current scattering off medium-heavy 
nuclei:$^{40}$Ca and $^{40}$Ar}
\author{A.~V.~Butkevich}
\affiliation{ Institute for Nuclear Research,
Russian Academy of Sciences,
60th October Anniversary Prosp. 7A,
Moscow 117312, Russia}
\date{\today}
\begin{abstract}

The charged-current quasi-elastic scattering of muon neutrinos on calcium and 
argon targets is calculated for neutrino energy up to 2.8 GeV. The calculations
 are done within the framework of the relativistic distorted-wave impulse 
approximation, which was earlier successfully applied to describe 
electron-nucleus data. The model is first tested against experimental data for 
electron scattering off calcium and then it is applied to calculate 
(anti)neutrino cross sections on ${}^{40}$Ca and ${}^{40}$Ar. We show that 
reduced exclusive cross sections for neutrino and electron scattering are 
similar. A significant nuclear model dependence of both inclusive and total 
cross sections for energy about 1 GeV was found. From the comparison of the 
(anti)neutrino differential and total cross sections per (proton)neutron, 
calculated for the carbon, oxygen, and argon targets it is evident that 
the cross sections decrease slowly with the mass-number of the target due to 
nuclear effects.
\end{abstract}

\maketitle

\section{Introduction}

The investigation of neutrino oscillations and the precision measurements of 
neutrino oscillation parameters brought extremely intense neutrino beams.
In this situation, statistical uncertainties are negligible compared to those
systematic uncertainties in the neutrino flux, neutrino-nucleus cross sections,
and detector effects on both neutrino events selection and neutrino energy 
reconstruction. Thus, the discovery and study of neutrino oscillations have 
renewed interest in neutrino-nucleus interactions. The high intensity neutrino 
beams allow to study these processes with unprecedented details.

In order to study neutrino oscillation effects in accelerator-based 
experiments, the neutrino beams cover the energy range from a few hundred MeV 
to several GeV. In this energy range, the dominant contribution to the 
neutrino-nucleus cross section comes from the charged-current quasielastic 
(CCQE) reaction and resonance production processes. The CCQE is the simplest 
interaction that represents a two-particle scattering process with single 
final-state proton and it may form a two track events sample. In this process 
the neutrino energy may be estimated using kinematic or calorimetric 
reconstruction.

The criteria used to select CCQE events is strongly influenced by both the 
target material and the detector technology. In fully active, fine-grained 
detectors with good resolution, selection techniques that relay on the 
identification of a single final state proton and lepton can be applied. One 
option for these detector designs is to make their active elements as 
water~\cite{SciFi}, scintillator~\cite{SciB,MINER,T2KN,NOVA} bars or drift 
chambers~\cite{NOMAD} formed in to detector planes. A continuous series of 
these planes forms a region that serves as both primary target and tracking 
detector.

Another option for fully active and fine-grained detectors is liquid argon 
time projection chamber (LAr TPC)~\cite{ICAR,Micro} which offers a relatively 
good particle identification. The high spatial resolution and energy 
measurement down to the MeV scale provides information for low and high energy 
particles. This information allows reduced background for the 
events of interest and potentially improved cross section measurements. The 
LAr TPC detectors are well-suited for long baseline $\nu_e$ appearance physics 
because of their high efficiency for $\nu_e$ 'signal' events and low 
background from $\nu_\mu$ events~\cite{LBNE}. Therefore, there is a growing 
interest in measuring neutrino cross sections on argon.
      
Unfortunately, the cross section data for lepton scattering on argon in relevant
 energy range are rather scarce. There are only experimental data for inclusive 
700-MeV electron scattering off ${}^{40}$Ar~\cite{Fras}. On the other hand the 
structures of ${}^{40}$Ca and ${}^{40}$Ar nuclei are almost identical and for 
calcium precise measurements have been performed and are described below.

High resolution exclusive $(e,e'p)$ experiments, where a proton is emitted with
 a direct knockout mechanism on ${}^{40}$Ca and ${}^{48}$Ca were carried out at 
Tokyo~\cite{CaT1,CaT2}, Saclay~\cite{CaS}, and NIKHEF~\cite{CaN1,CaN2,CaN3}. 
Specific quantum numbers and spectroscopic factors have been assigned to the 
peaks in the observed energy spectrum by studying the missing energy and 
momentum dependence of experimental cross sections. The data analysis of 
this processes was performed within the theoretical framework of the 
nonrelativistic distorted-wave impulse approximation (DWIA)~\cite{Gi1,Gi2} and 
relativistic DWIA (RDWIA)~\cite{Kelly} in 
Refs.~\cite{CaN1,CaN2,CaN3,Lapik,McD,Jin,Udi1,Udi2,Giusti}. 
There, the RDWIA approach was able to describe with high degree of accuracy 
the experimental shape of the outgoing particle momentum distributions.  
In order to reproduce experimental cross sections, normalization of the 
bound-state wave functions were fitted to the data and identified with the 
spectroscopic factors.

Inclusive $(e,e')$ cross sections of the electron scattering on calcium were 
measured with good accuracy in a series of 
experiments~\cite{Whit,Mez,Dead,Will,Yat}. In Refs.~\cite{Dead,Will} the 
transverse and longitudinal nuclear response functions were extracted. The 
comparison of different models with the data is described in 
Refs~\cite{BAV1,Sob}.

In this work we compute the exclusive, inclusive and total cross sections for 
the CCQE neutrino scattering from ${}^{40}$Ca and ${}^{40}$Ar using the RDWIA 
model. The calculations of the inclusive cross sections are performed with our
 approach, which includes the final state interaction (FSI) effects in the 
presence of short-range nucleon-nucleon ($NN$) correlations in the ground 
state~\cite{BAV2}. This approach was successfully applied in Refs.
~\cite{BAV3,BAV4,BAV5,BAV6}. First we compare our model in 
describing ${}^{40}$Ca$(e,e'p)^{39}$K and ${}^{40}$Ca$(e,e')$ data. 
Then we apply it to the calculation of the CCQE cross sections for the 
neutrino scattering on ${}^{40}$Ca and ${}^{40}$Ar nuclei. 

The goals of this work are the following: (a) calculation of the RDWIA CCQE 
$\nu^{40}$Ar cross sections, (b) comparison of the total cross sections, scaled with the number of (proton)neutrons in the target for (anti)neutrino scattering on the carbon, oxygen, and argon targets, and (c) investigation of nuclear 
effects on these cross sections.

The paper is organized as follows. In Sec. II we present briefly the formalism 
for the CCQE scattering process and basic aspects of the model used for the 
calculation. The results are presented and discussed in Sec. III. Our 
conclusions are summarized in Sec. IV.  

\section{Formalism of quasi-elastic scattering and RDWIA}

In this section we consider shortly the formalism used to describe electron 
and neutrino quasi-elastic exclusive
\begin{equation}\label{qe:excl}
l(k_i) + A(p_A)  \rightarrow l^{\prime}(k_f) + N(p_x) + B(p_B),      
\end{equation}
and inclusive
\begin{equation}\label{qe:incl}
l(k_i) + A(p_A)  \rightarrow l^{\prime}(k_f) + X                      
\end{equation}
scattering off nuclei in the one-photon (W-boson) exchange approximation. 
Here $l$ labels the incident lepton [electron or muon (anti)neutrino], and
$l^{\prime}$ represents the scattered lepton (electron or muon),
$k_i=(\varepsilon_i,\k_i)$ 
and $k_f=(\varepsilon_f,\k_f)$ are the initial and final lepton 
momenta, $p_A=(\varepsilon_A,\p_A)$, and $p_B=(\varepsilon_B,\p_B)$ are 
the initial and final target momenta, $p_x=(\varepsilon_x,\p_x)$ is the 
ejectile nucleon momentum, $q=(\omega,\q)$ is the momentum transfer carried by 
the virtual photon (W-boson), and $Q^2=-q^2=\q^2-\omega^2$ is the photon 
(W-boson) virtuality. 

\subsection{ CCQE neutrino-nucleus cross sections}

In the laboratory frame, the differential cross section for the exclusive
electron ($\sigma ^{el}$) and (anti)neutrino CCQE ($\sigma ^{cc}$) scattering, 
in which only a single discrete state or narrow resonance of the target is 
excited, can be written as
\begin{subequations}
\label{cs5}
\begin{align}
\frac{d^5\sigma^{el}}{d\varepsilon_f d\Omega_f d\Omega_x} &= R
\frac{\vert\p_x\vert{\varepsilon}_x}{(2\pi)^3}\frac{\varepsilon_f}
{\varepsilon_i} \frac{\alpha^2}{Q^4} L_{\mu \nu}^{(el)}W^{\mu \nu (el)}
\\                                                                       
\frac{d^5\sigma^{cc}}{d\varepsilon_f d\Omega_f d\Omega_x} &= R
\frac{\vert\p_x\vert{\varepsilon}_x}{(2\pi)^5}\frac{\vert\k_f\vert}
{\varepsilon_i} \frac{G^2\cos^2\theta_c}{2} L_{\mu \nu}^{(cc)}W^{\mu \nu (cc)},
\end{align}
\end{subequations}
 where $R$ is a recoil factor, $\Omega_f$ is the solid angle for the lepton 
momentum, $\Omega_x$ is the solid angle for the ejectile nucleon momentum, 
$\alpha\simeq 1/137$ is the fine-structure constant, $G \simeq$ 1.16639 
$\times 10^{-11}$ MeV$^{-2}$ is the Fermi constant, $\theta_C$ is the Cabbibo 
angle ($\cos \theta_C \approx$ 0.9749), $L_{\mu \nu}$ is the lepton tensor and 
$W^{(el)}_{\mu \nu}$ and $W^{(cc)}_{\mu \nu}$ are the electromagnetic and weak CC 
nuclear tensors, respectively. The energy $\varepsilon_x$ is the solution to 
the equation
\begin{equation}\label{eps}
\varepsilon_x+\varepsilon_B-m_A-\omega=0,                                 
\end{equation}
where $\varepsilon_B=\sqrt{m^2_B+\p^2_B}$, $~\p_B=\q-\p_x$, $~\p_x=
\sqrt{\varepsilon^2_x-m^2}$, and $m_A$, $m_B$, and $m$ are masses of the 
target, recoil nucleus and nucleon, respectively. 
The missing momentum $p_m$ and missing energy $\varepsilon_m$ are defined by 
\begin{subequations}
\begin{align}
\label{p_m}
\p_m & = \p_x-\q
\\
\label{eps_m}
\varepsilon_m & = m + m_B - m_A.                                          
\end{align}
\end{subequations}

The leptonic tensor is separated into a symmetrical and an anti-symmetrical
components that are written as in Ref.~\cite{BAV2}. For electron scattering 
at energies considered in this work, a simple and accurate enough effective 
momentum approximation~\cite{Boffi,Udi1} to include the effects of Coulomb 
distorted wave functions is used. 
All information about the nuclear structure and FSI effects is 
contained in the electromagnetic and weak CC hadronic tensors,
$W^{(el)}_{\mu \nu}$ and $W^{(cc)}_{\mu \nu}$, which are given by the bilinear 
products of the transition matrix elements of the nuclear electromagnetic or CC 
operator $J^{(el)(cc)}_{\mu}$ between the initial nucleus state 
$\vert A \rangle $ and the final state $\vert B_f \rangle$ as
\begin{eqnarray}
W^{(el)(cc)}_{\mu \nu } &=& \sum_f \langle B_f,p_x\vert                     
J^{(el)(cc)}_{\mu}\vert A\rangle \langle A\vert
J^{(el)(cc)\dagger}_{\nu}\vert B_f,p_x\rangle,              
\label{W}
\end{eqnarray}
where the sum is taken over undetected states.

The experimental data of the $(e,e'p)$ reaction are usually presented in terms 
of the reduced cross section
\begin{equation}
\sigma_{red} = \frac{d^5\sigma}{d\varepsilon_f d\Omega_f d\Omega_x}
/K^{(el)(cc)}\sigma_{lN},                                                 
\end{equation}
where
$K^{el} = R {p_x\varepsilon_x}/{(2\pi)^3}$ and
$K^{cc}=R {p_x\varepsilon_x}/{(2\pi)^5}$
are phase-space factors for electron and neutrino scattering  and 
$\sigma_{lN}$ is the corresponding
elementary cross section for the lepton scattering from the moving free
nucleon. The reduced cross section is an interesting quantity that can be 
regarded as the nucleon momentum distribution modified by FSI. It was shown in
Refs.~\cite{BAV2,BAV4} that reduced cross sections for (anti)neutrino 
scattering off carbon and oxygen are similar to the electron scattering apart 
from small differences at low beam energy due to effects of Coulomb 
distortion of the incoming electron wave function.

\subsection{ Models}

We describe the lepton-nuclear scattering in the impulse 
approximation (IA), assuming that the incoming lepton interacts with only one 
nucleon of the target, which is subsequently emitted, while the remaining 
(A-1) nucleons in the target are spectators. When the nuclear current is 
written as a sum of single-nucleon currents, the nuclear matrix element 
in Eq.(\ref{W}) can write as
\begin{eqnarray}\label{Eq12}
\langle p,B\vert J^{\mu}\vert A\rangle &=& \int d^3r~ \exp(i\t\cdot\r)
\overline{\Psi}^{(-)}(\p,\r)
\Gamma^{\mu}\Phi(\r),                                                   
\end{eqnarray}
where $\Gamma^{\mu}$ is the vertex function, $\t=\varepsilon_B\q/W$ is the
recoil-corrected momentum transfer, $W=\sqrt{(m_A + \omega)^2 - \q^2}$ is the
invariant mass, $\Phi$ and $\Psi^{(-)}$ are the relativistic bound-state and
outgoing wave functions.

For electron scattering, we use the CC2 electromagnetic vertex
function for a free nucleon~\cite{deFor}
\begin{equation}
\Gamma^{\mu} = F^{(el)}_V(Q^2)\gamma^{\mu} + {i}\sigma^{\mu \nu}\frac{q_{\nu}}
{2m}F^{(el)}_M(Q^2),                                                    
\end{equation}
where $\sigma^{\mu \nu}=i[\gamma^{\mu},\gamma^{\nu}]/2$, $F^{(el)}_V$ and
$F^{(el)}_M$ are the Dirac and Pauli nucleon form factors. 
The single-nucleon charged current has $V{-}A$ structure $J^{\mu(cc)} = 
J^{\mu}_V + J^{\mu}_A$. For a free-nucleon vertex function 
$\Gamma^{\mu(cc)} = \Gamma^{\mu}_V + \Gamma^{\mu}_A$ we use the CC2 vector 
current vertex function
\begin{equation}
\Gamma^{\mu}_V = F_V(Q^2)\gamma^{\mu} + {i}\sigma^{\mu \nu}\frac{q_{\nu}}
{2m}F_M(Q^2)                                                          
\end{equation}
and the axial current vertex function
\begin{equation}
\Gamma^{\mu}_A = F_A(Q^2)\gamma^{\mu}\gamma_5 + F_P(Q^2)q^{\mu}\gamma_5. 
\end{equation}
The weak vector form factors $F_V$ and $F_M$ are related to the corresponding 
electromagnetic form factors $F^{(el)}_V$ and $F^{(el)}_M$ for protons and 
neutrons by the hypothesis of the conserved vector current. We use the 
approximation of Ref.~\cite{MMD} for the Dirac and Pauli nucleon form factors. 
Because the bound nucleons are the off-shell we employ the de Forest 
prescription~\cite{deFor} and Coulomb gauge for the off-shell vector current 
vertex $\Gamma^{\mu}_V$. The vector-axial $F_A$ and pseudoscalar $F_P$ form 
factors are parametrized using a dipole approximation: 
\begin{equation}
F_A(Q^2)=\frac{F_A(0)}{(1+Q^2/M_A^2)^2},\quad                         
F_P(Q^2)=\frac{2m F_A(Q^2)}{m_{\pi}^2+Q^2},
\end{equation}
where $F_A(0)=1.267$, $M_A$ is the axial mass, which controls $Q^2$-dependence 
of $F_A(Q^2)$, and $m_\pi$ is the pion mass.

In RDWIA calculations the independent particle shell model (IPSM) is assumed 
for the nuclear structure. In Eq.(\ref{Eq12}) the relativistic bound-state wave 
function for nucleons $\Phi$ are four-spinors that are obtained as the 
self-consistent (Hartree--Bogolioubov) solutions of a Dirac equation, derived 
within a relativistic mean-field approach, from a Lagrangian containing 
$\sigma$, $\omega$, and $\rho$ mesons~\cite{Hor1,Hor2,Hor3}. We use the nucleon 
bound-state functions calculated by the TIMORA code~\cite{Hor3} with the
normalization factors $S(\alpha)$ relative to full occupancy of the IPSM 
orbitals of ${}^{40}$Ca and ${}^{40}$Ar. The source of the reduction of the 
$(e,e'p)$ spectroscopic factors with respect to the mean field values are 
the short-range and tensor correlations, which arise from the characteristics 
of the bare nucleon-nucleon interaction and long-range correlations related to 
the coupling between single-particle motion and collective surface vibrations. 

For an outgoing nucleon, the simplest choice is to use the plane-wave function 
$\Psi$, assuming that there is no interaction between the ejected nucleon $N$
and the residual nucleus $B$. In this plane-wave impulse approximation (PWIA)
 the exclusive cross section is factorized into the product of the phase-space 
factor $K$, elementary off-shell lepton-nucleon scattering cross section, and 
the hole spectral function. Thus, in the PWIA the reduced cross section can be 
interpreted as the momentum distribution of the emitted nucleon when it was 
inside the nucleus. For more realistic description, FSI effects 
should be taking into account. In the RDWIA the distorted wave function $\Psi$
 is evaluated as a solution of the Dirac equation containing a 
phenomenological relativistic optical potential. This potential consists of a 
real part, which describes the rescattering of the ejected nucleon and an 
imaginary part which accounts for its absorption into unobserved channels.

Using the direct Pauli reduction method~\cite{Udi2}, the system of two coupled 
first-order Dirac equations can be reduced to a single second-order 
Schr\"odinger-like equation for the upper component of the Dirac wave function 
$\Psi$. This equation contains equivalent nonrelativistic central and 
spin-orbit potentials which are functions of the relativistic, energy 
dependent, scalar, and vector optical potentials. 
We use the LEA program~\cite{LEA} for the numerical calculation of the 
distorted wave functions with the EDAD1 parametrization~\cite{Cooper} of the 
relativistic optical potential for calcium. This code was successfully tested 
against $A(e,e'p)$ data for electron scattering off ${}^{12}$C and ${}^{16}$O
~\cite{Kelly2,Fissum} and we adopted this program for neutrino 
reaction~\cite{BAV2}.   

A complex optical potential with a nonzero imaginary part generally produces 
an absorption of the flux. For the exclusive $A(l,l'N)$ channel this reflects 
the coupling between different open reaction channels. However, for the 
inclusive reaction, the total flux must be conserved. 
In Refs.~\cite{Meu1,Meu2}, it was shown that the inclusive CCQE neutrino 
cross section of the exclusive channel $A(l,l'N)$ is calculated with only the 
real part of the optical potential is almost identical when calculated via the 
Green's function approach~\cite{Meu1}, in which the FSI effects on 
inclusive reaction $A(l,l'X)$ are treated by means of a complex potential, and 
the total flux is conserved. We calculate the inclusive and total cross 
sections with the EDAD1 relativistic optical potential in which only the real 
part is included. The inclusive cross sections with the FSI effects 
in the presence of the short-range $NN$ correlations are calculated using the 
method proposed in Ref.~\cite{BAV2}. In this approach the contribution of the 
$NN$ correlated pairs is evaluated in the PWIA. The FSI effects for the 
high-momentum component are estimated by scaling the PWIA cross section with 
the $\Lambda(\varepsilon_f,\Omega_f)$ function determined in Ref.~\cite{BAV2}. 

\section{Results and analysis}

\subsection{Electron scattering}

In this work the IPSM is assumed for the ${}^{40}$Ca and ${}^{40}$Ar nuclear 
structures. 
The model space for ${}^{40}$Ca$(l,l^{\prime}N)$ consists of
1$s_{1/2}$, 1$p_{3/2}$, 1$p_{1/2}$, 1$d_{5/2}$, 2$s_{1/2}$ and 1$d_{3/2}$ 
nucleon-hole states in ${}^{39}$K and ${}^{39}$Ca nuclei. 
\begin{figure*}
  \begin{center}
    \includegraphics[height=16cm,width=16cm]{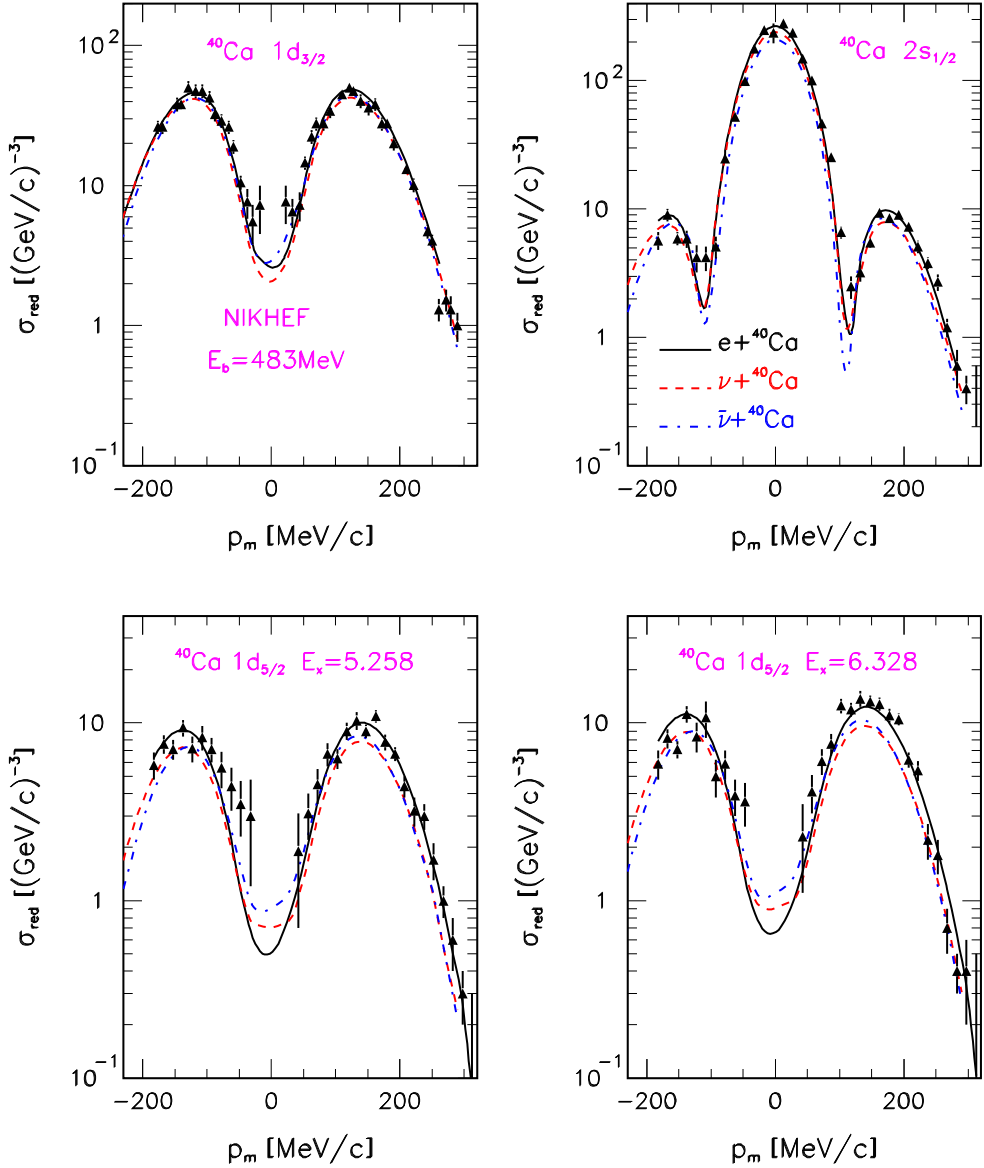}
  \end{center}
  \caption{(Color online) Comparison of the RDWIA calculations for electron 
(solid line), neutrino (dashed line), and antineutrino (dashed-dotted line) 
reduced  cross sections for the removal of nucleons from 
1$d_{3/2}$, 2$s_{1/2}$, and 1$d_{5/2}$ shells of $^{40}$Ca with NIKHEF 
data~\cite{CaN3}. We show the results obtained in $(\q,\omega)$ constant 
kinematics for electron  beam energy $E_{beam}$=483.2 MeV, the outgoing proton 
kinetic energy $T_p$=100 MeV, and $\q$=450 (MeV/c)$^2$. The cross sections are 
presented as functions of missing momentum $p_m$ for the transition to the 
3/2$^{+}$ ground state, 1/2$^{+}$ ($E_x$=2.522 MeV), and 5/2$^{+}$ ($E_x$=5.258, 
6.328 MeV) excited states of ${}^{39}$K and ${}^{39}$Ca.} 
\end{figure*}
The model space for ${}^{40}$Ar$(l,l^{\prime}N)$ consists of  
1$s_{1/2}$, 1$p_{3/2}$, 1$p_{1/2}$, 1$d_{5/2}$, 2$s_{1/2}$ and 1$d_{3/2}$ 
nucleon-hole states in ${}^{39}$Cl, and 1$s_{1/2}$, 1$p_{3/2}$, 1$p_{1/2}$, 
1$d_{5/2}$, 2$s_{1/2}$, 1$d_{3/2}$ and 1$f_{7/2}$ nucleon-hole states in 
${}^{39}$Ar. All states are regarded as a discrete states even though their 
spreading widths are actually appreciable.
\begin{figure*}
  \begin{center}
    \includegraphics[height=16cm,width=16cm]{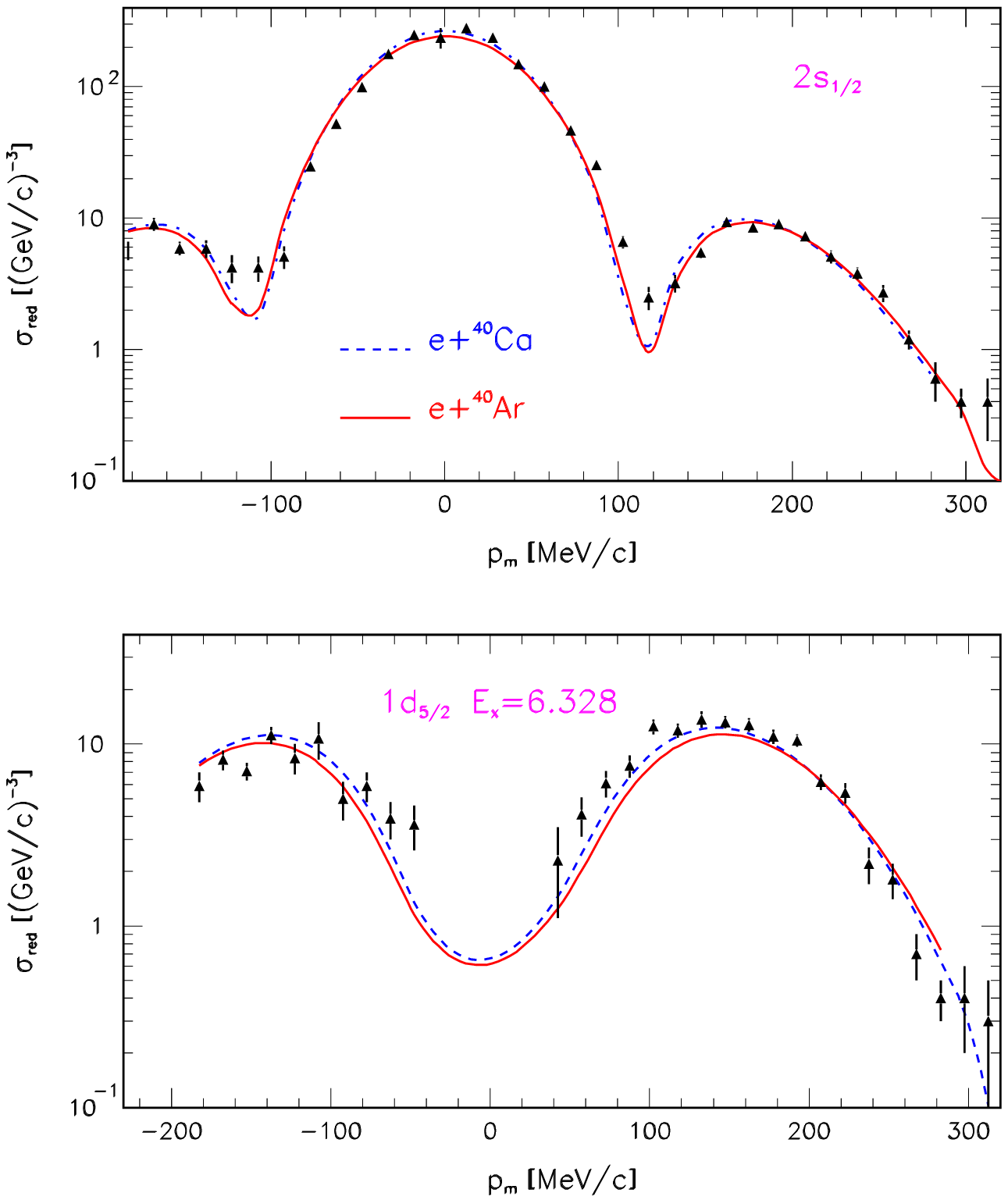}
  \end{center}
  \caption{(Color online) Reduced cross sections of the $^{40}$Ar$(e,e'p)$ 
(solid line) and $^{40}$Ca$(e,e'p)$ (dashed line) reactions as functions of  
missing momentum $p_m$ for the transitions to the 1/2$^{+}$ ($E_x$=2.522 MeV) 
and 5/2$^{+}$ ($E_x$=6.328 MeV) excited states in $^{39}$Cl and $^{39}$K. For 
comparison the data for the $^{40}$Ca$(e,e'p)$ reaction are shown from 
Ref.~\cite{CaN3}.}   
\end{figure*}

First we show the performances of the LEA program in describing of 
experimental data for the ${}^{40}$Ca$(e,e^{\prime}p)$ reaction measured at 
NIKHEF in $(\q,\omega)$ constant kinematics~\cite {CaN1,CaN3}. The comparison 
with the experimental reduced cross sections is displayed in Fig.1. We have 
considered cross sections for the removal of the proton from the 1$d_{3/2}$ 
shell, for transition to the 1/2$^{+}$ excited state of the ${}^{39}$K nucleus 
at the excitation energy of $E_x$=2.522 MeV, and for the transitions to the 
5/2$^{+}$ excited states at $E_x$=5.258 MeV and $E_x$=6.328 MeV, obtained by knocking out
 protons from the 2$s_{1/2}$ and 1$d_{5/2}$ orbitals, correspondingly. It should 
be noted that positive (negative) values of $p_m$ refer to situations where the 
angle between the outgoing proton momentum $\p_m$ and the incident electron 
$\k_i$ is larger (smaller) than the angle between $\q$ and $\k_i$.

The missing momentum distribution calculated in the RDWIA approach is 
shown in Fig.1 with NIKHEF data~\cite{CaN1} and provide a good description of 
the shape of the measured distribution. Normalization factors have 
been applied to reproduce the magnitude of the measured reduced cross sections.
 The factors of 0.68, 0.51, 0.11, and 0.13 for the transition to the 3/2$^{+}$,
 1/2$^{+}$ ($E_x$=2.522 MeV), 5/2$^{+}$ ($E_x$=5.258 MeV), and 5/2$^{+}$ 
($E_x$=6.328 MeV) states, correspondingly are almost identical to those 
obtained in the data analysis of Refs.~\cite{CaN3,Giusti}. Neutrino and 
antineutrino reduced cross sections of ${}^{40}$Ca$(\nu,\mu^{-}p){}^{39}$Ca and 
${}^{40}$Ca$(\bar{\nu},\mu^{+}n){}^{39}$K reactions also shown are in Fig.1. 
They were calculated with the same reduced factors as electron cross sections. 
There is an overall good agreement between calculated cross sections,
but the values of the electron cross sections at the maximum is systematically
higher than those for (anti)neutrino. This can be attributed to Coulomb 
distortion upon the incident electron wave function. The small difference 
between neutrino and antineutrino is due to difference in the FSI of the 
proton and neutron with the residual nucleus.

In Fig.2 we compare our results for the ${}^{40}$Ar$(e,e^{\prime}p)$ reaction for 
transitions to the 1/2$^{+}$ ($E_x$=2.522 MeV) and 5/2$^{+}$ ($E_x$=6.328 MeV) 
excited states of ${}^{39}$K with the ${}^{40}$Ca$(e,e^{\prime}p)$ data. Also in 
this case, the results are multiplied by the same normalization factors (0.51 
and 0.13) as for the ${}^{40}$Ca$(e,e^{\prime}p)$ reduced cross sections. In this 
figure for comparison shown are the calculated reduced cross-section for the 
${}^{40}$Ca$(e,e^{\prime}p)$ reaction. 
The cross sections for the removal protons from the 1/2$^{+}$ and 5/2$^{+}$ 
shells of ${}^{40}$Ca and ${}^{40}$Ar as functions of $p_m$ are very similar, 
but at the maximum the values of the cross sections for ${}^{40}$Ar are 
systematically lower (less than 12\%) than for ${}^{40}$Ca.
%
\begin{table}
\begin{ruledtabular}
\caption{\small Proton and neutron binding energies ($E_b$) and the 
occupancies for ${}^{40}$Ca.}
\begin{tabular}{lcccc}
Orbital & \multicolumn{2}{c}{     $E_b$ (MeV)} &  S 
 \\[1mm]
 &   p & n & & 
 \\[1mm]
 \hline 
1s$_{1/2}$ & 57.4 &    64.3 & 1  &     \\[1mm]
1p$_{3/2}$ & 36.5    & 43.5 & 0.95  &   \\[1mm]
1p$_{1/2}$ & 31.6    & 38.7 & 0.95  &   \\[1mm]
1d$_{5/2}$ & 15.4    & 22.6 & 0.80  &   \\[1mm]
2s$_{1/2}$ & 10.9    & 18.1 & 0.85  &   \\[1mm]
1d$_{3/2}$ & 8.3    & 15.6 & 0.82  &   \\[1mm]
\end{tabular}
\label{tab:Ca1}
\end{ruledtabular}
\end{table}

Mean values of the proton and neutron binding energies and occupancies of 
shells that used in this work are listed in Table I for ${}^{40}$Ca and in 
Table II for ${}^{40}$Ar. The values of the proton and neutron binding energies
 for the 1$d_{3/2}$, 2$s_{1/2}$, and 1$d_{5/2}$ shells, as well as those for 
neutron for the 1$f_{7/2}$ orbital in ${}^{40}$Ar were taken from 
Ref.~\cite{BNL}. For the 1$p_{3/2}$, 1$p_{1/2}$, and 1$s_{1/2}$ deeply bound 
orbitals proton and neutron binding energies were estimated in Refs.
~\cite{Tornow,Mah}. In Ref.~\cite{CaN3} the 2$p_{3/2}$, 1$f_{7/2}$, 1$d_{3/2}$, 
2$s_{1/2}$, and 1$d_{5/2}$ strengths were obtained as a sum of the strengths 
arising from the discrete transitions and strengths which are observed in the 
continuum at higher excitation energies. In Table I occupancy of the 
1$d_{3/2}$, 2$s_{1/2}$, and 1$d_{5/2}$ orbitals from Ref.~\cite{CaN3} are shown. 
The occupancy of the 1$p_{3/2}$ and 1$p_{1/2}$ shells were estimated from the 
${}^{40}$Ca$(e,e^{\prime}p)$ data analysis in Ref.~\cite{CaS}. For the 1$s_{1/2}$ 
shell we assume that occupancy is equal to 1.
\begin{table}
\begin{ruledtabular}
\caption{\small Proton and neutron binding energies ($E_b$) and the 
occupancies for ${}^{40}$Ar.}
\begin{tabular}{lccccc}
Orbital & \multicolumn{2}{c}{     $E_b$ (MeV)} &\multicolumn{2}{c}{S} 
 \\[1mm]
 &   p & n & p & n & 
 \\[1mm]
 \hline 
1s$_{1/2}$ & 57.4 &    64.3 & 1  &  1  &   \\[1mm]
1p$_{3/2}$ & 36.5    & 43.5 & 0.95  &  0.95 & \\[1mm]
1p$_{1/2}$ & 31.6    & 38.7 & 0.95  &  0.95  \\[1mm]
1d$_{5/2}$ & 15.4    & 22.6 & 0.80  &  0.80  \\[1mm]
2s$_{1/2}$ & 10.9    & 18.1 & 0.85  &  0.85  \\[1mm]
1d$_{3/2}$ & 8.3    & 15.6  & 0.85  &   0.82  \\[1mm]
1f$_{7/2}$ &        & 9.87  &       &   0.82  \\[1mm]
\end{tabular}
\label{tab:Ca2}
\end{ruledtabular}
\end{table}
 
Note, that in the IPSM the 20 protons and 20 neutrons of ${}^{40}$Ca fill the 
shells up to the 1$d_{3/2}$ orbital whereas the first empty orbital is the 
1$f_{7/2}$ orbital. Ground-state correlations manifest themselves by a depletion
 of the orbitals bellow and a filling the orbitals above Fermi level, i.e. 
nucleons from the 1$d_{3/2}$ and 2$s_{1/2}$ shells are promoted to the 1$f_{7/2}$ 
and 2$s_{1/2}$ orbitals. We include the observed 1$f_{7/2}$ ($E_x$=2.814 MeV) and 
2$p_{3/2}$ ($E_x$=3.019 MeV) strength of 0.36 in the 2$s_{1/2}$ shell because 
the observed missing-energy spectrum~\cite{CaN3} of these orbitals strongly 
overlap. The rest of the 1$f_{7/2}$ and 2$p_{3/2}$ strength were included in the
 1$d_{3/2}$ shell. In the IPSM the 18 protons (22 neutrons) of ${}^{40}$Ar fill 
the shells up to the 1$d_{3/2}$ (1$f_{7/2}$) shell and the occupancies of the 
shells are not measured. For ${}^{40}$Ar we assume the same occupancies of the 
orbitals as for ${}^{40}$Ca because the structures of these nuclei are almost 
identical.

In our approach the occupancy of the IPSM orbitals of ${}^{40}$Ca and 
${}^{40}$Ar are approximately 87\% on average. We assume that the missing 
strength can be attributed to the short-range $NN$ correlations, leading to 
the appearance of the high-momentum and high-energy nucleon distribution in 
the target. We use the general expression~\cite{Kozel} for the high-momentum 
part of the spectral function $P_{HM}$ with the parametrization of the momentum 
distribution for ${}^{40}$Ca taken from Ref.~\cite{Ciofi}. In our calculations, 
the spectral function $P_{HM}$ incorporates 13\% of the IPSM sum-rile limit.
\begin{figure*}
  \begin{center}
    \includegraphics[height=16cm,width=16cm]{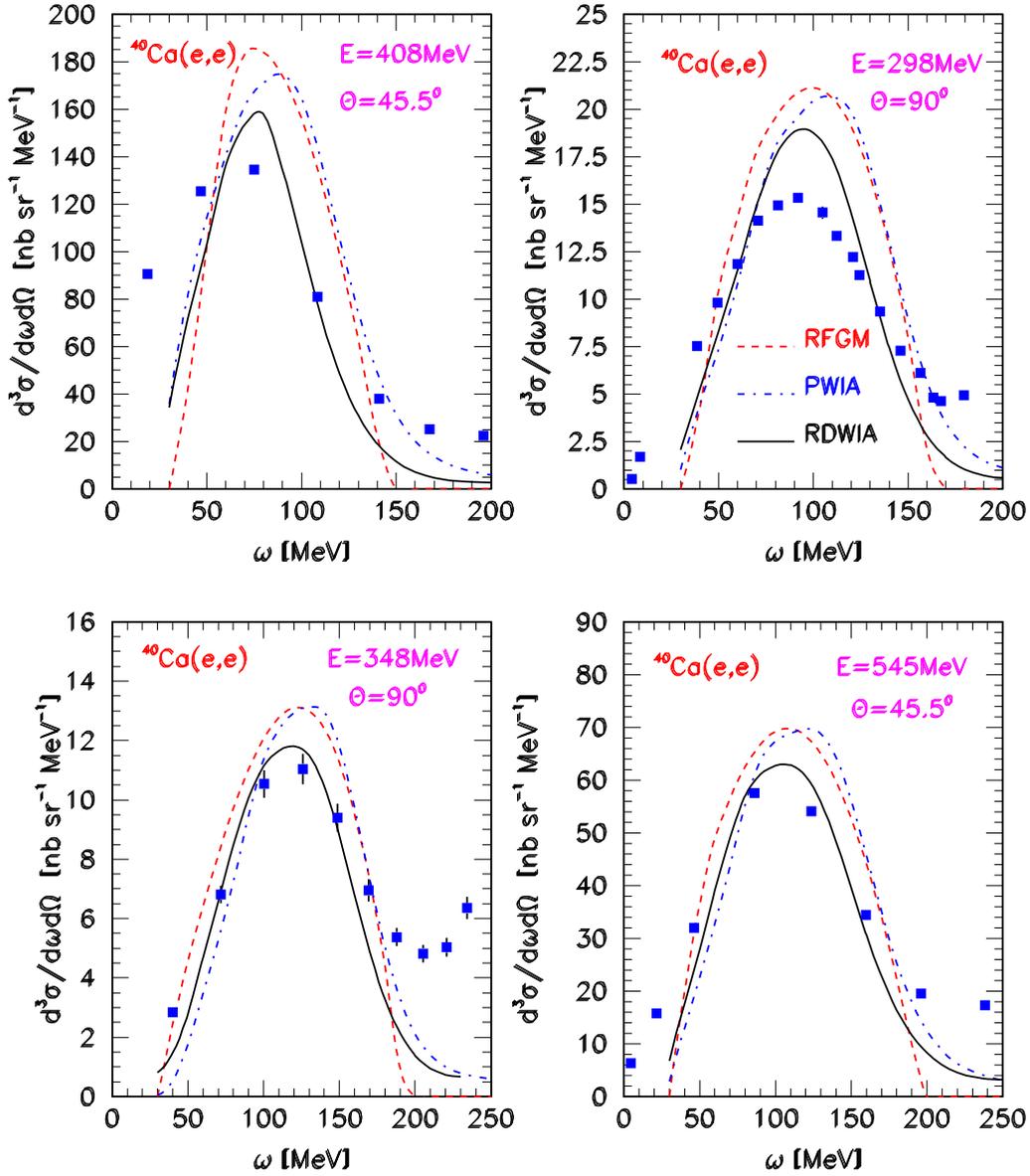}
  \end{center}
  \caption{(Color online) Inclusive cross section versus energy transfer
$\omega$ for electron scattering on $^{40}$Ca. The data are from 
Ref.\cite{Will} for the electron beam energies $E_e$=408, 545 MeV and 
scattering angle $\theta$=45.5$^{\circ}$; $E_e$=298, 348 MeV and 
$\theta$=90$^{\circ}$. As shown in the the key, cross sections were calculated 
with the RDWIA, PWIA, and RFGM.}
\end{figure*}
\begin{figure}
  \begin{center}
    \includegraphics[height=16cm,width=16cm]{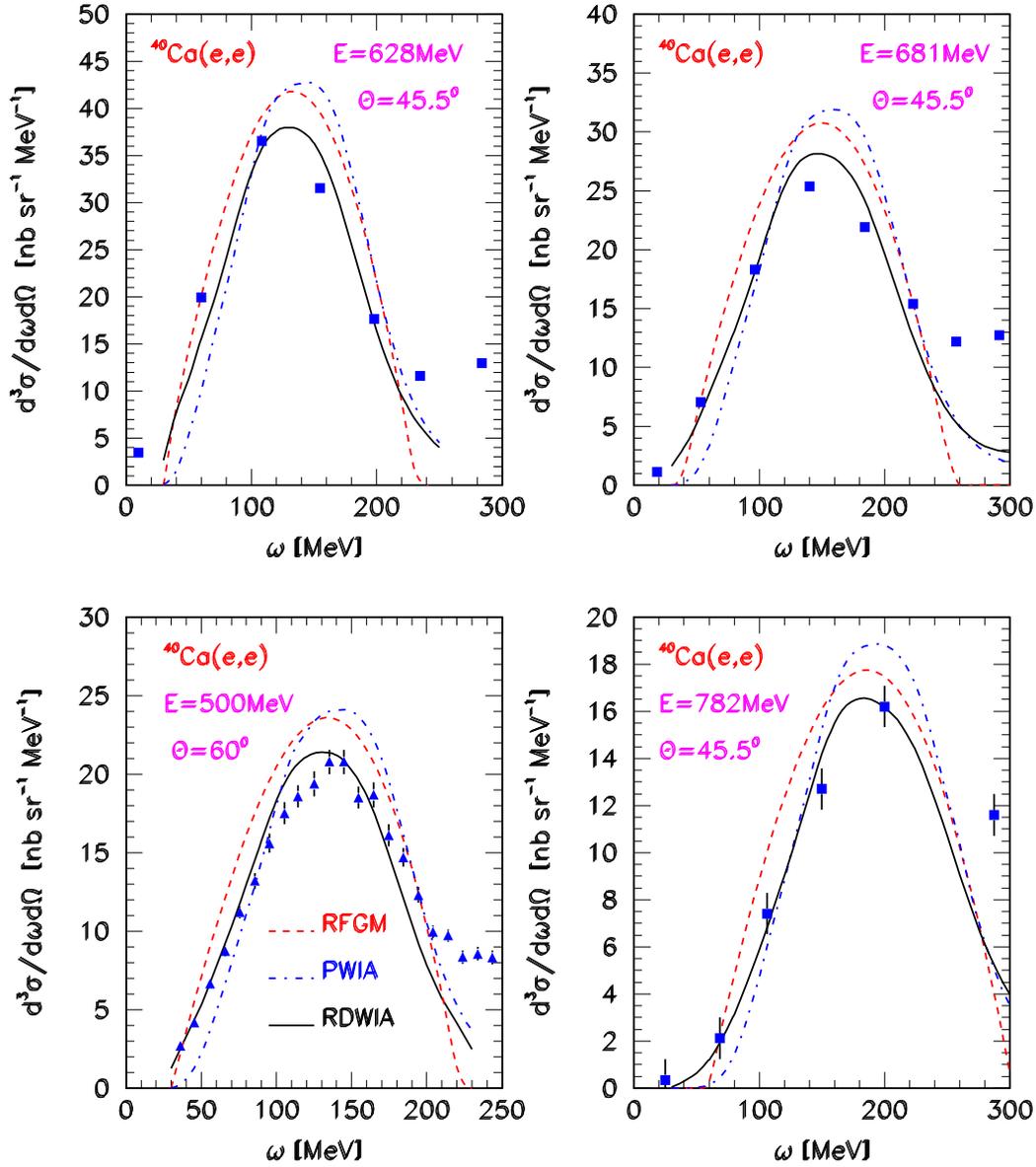}
  \end{center}
  \caption{(Color online) Same as Fig. 3, but the data are from Ref.~\cite{Will}
    (squares) for $E_e$=628, 681, 782 MeV, $\theta$=45.5$^{\circ}$ and from 
Ref.\cite{Whit} (triangles) for $E_e$=500 MeV, $\theta$=60$^{\circ}$.}   
\end{figure}
The inclusive cross sections with the FSI effects in the presence of the 
short-range $NN$ correlations were calculated using the approach proposed 
in Ref.~\cite{BAV2}. To test our approach, we calculated the inclusive 
${}^{40}$Ca$(e,e^{\prime})$ cross sections and compared them with data from SLAC
~\cite{Whit} and Bates~\cite{Will} experiments. Figures 3 and 4 show measured 
inclusive cross sections as functions of energy transfer as compared to the 
RDWIA, PWIA, and relativistic Fermi gas model (RFGM) calculations with the 
Fermi momentum $p_F$=249 MeV and nuclear binding energy $\epsilon_b$=33 MeV. 
These data cover the range of the three-momentum transfer (around the peak) 
from $|\q|\approx$300 MeV/c (beam energy $E_e$=408 MeV and scattering angle 
$\theta=45.5^{\circ}$ up to $|\q|\approx$560 MeV/c ($E_e$=782 MeV,
$\theta=45.5^{\circ}$). We note that relative to the PWIA results, the generic 
effect of the FSI reduces the cross section value around the QE peak and 
shifts the peak toward lower values of the energy transfer. The peak in the 
RDWIA calculation occurs at the same energy loss as the data and the value of 
the calculated cross sections (apart from $E_e$=298 MeV, $\theta=90^{\circ}$) 
generally agree with data within 14\%. On the other hand the PWIA and RFGM 
results systematically overestimate the data. 
 
\subsection{Neutrino scattering}

To study nuclear effects on the $Q^2$ distribution, we calculated with
$M_A$=1.032 GeV the inclusive cross sections $d\sigma/dQ^2$ for
(anti)neutrino energies $\varepsilon_{\nu}$=0.5, 0.7, 1.2 and 2.5 GeV in the 
RDWIA and RFGM approaches. 
\begin{figure*}
  \begin{center}
    \includegraphics[height=16cm,width=16cm]{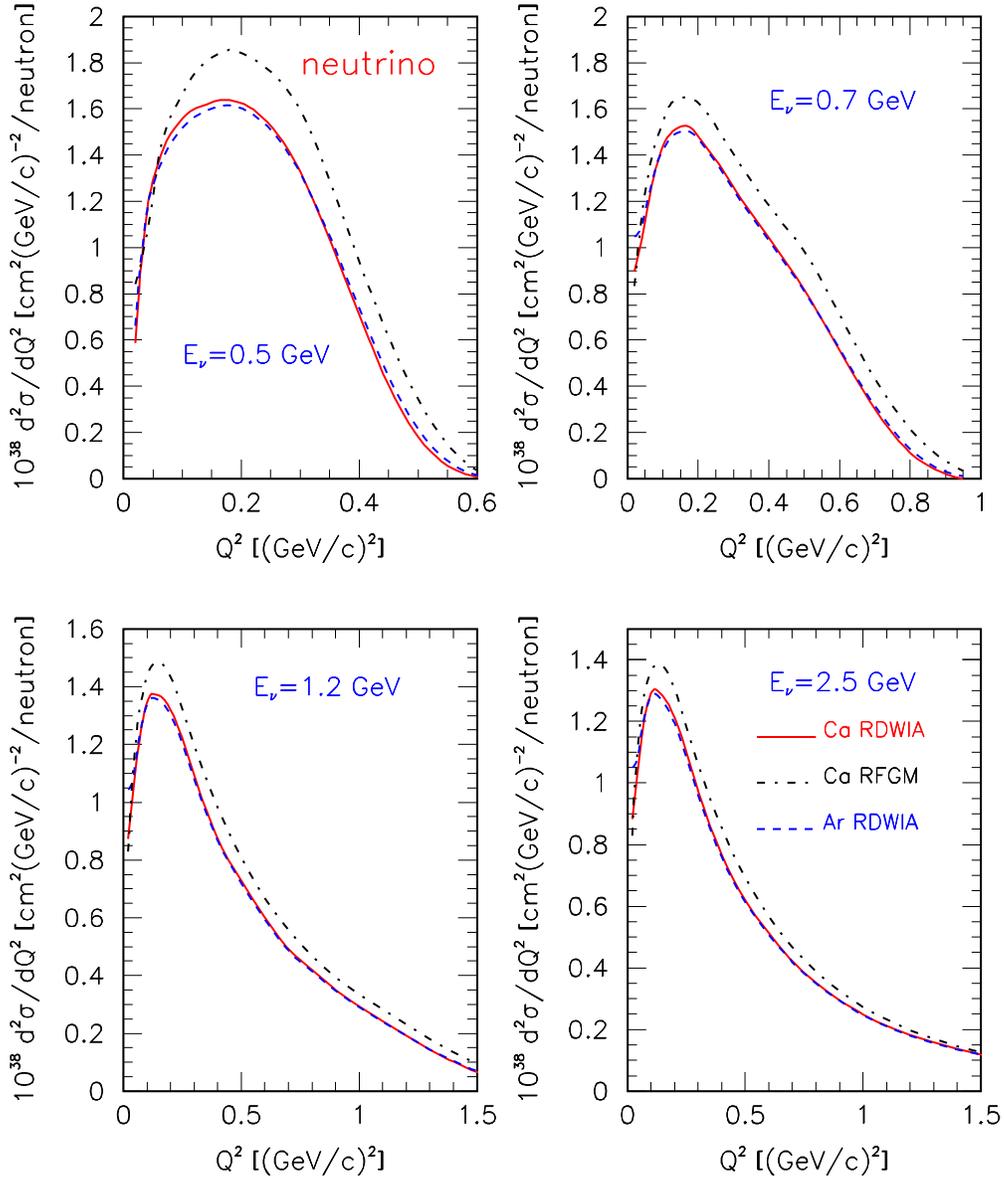}
  \end{center}
  \caption{(Color online) Inclusive cross section per neutron vs. the 
four-momentum transfer $Q^2$ for neutrino scattering on ${}^{40}$Ca and 
${}^{40}$Ar  and for the four values of incoming neutrino energy:
$\varepsilon_{\nu}$=0.5, 0.7, 1.2, and 2.5 GeV. As shown in the key, the cross 
sections were calculated with the RDWIA and RFGM (for calcium) approaches.}
\end{figure*}
\begin{figure*}
  \begin{center}
    \includegraphics[height=16cm,width=16cm]{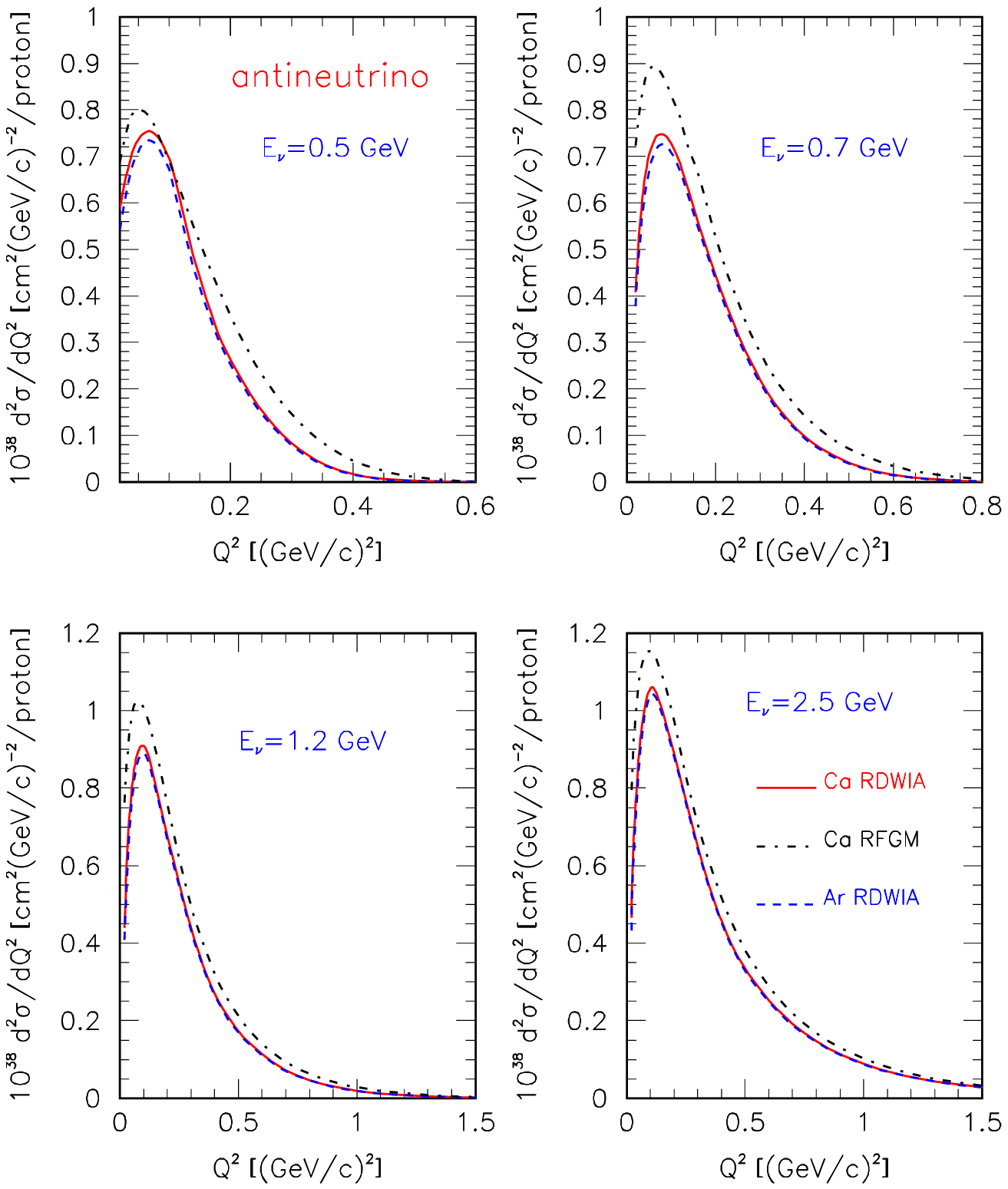}
  \end{center}
  \caption{(Color online) Same as Fig. 5, but cross section per proton for 
antineutrino scattering.} 
\end{figure*}
The results for neutrino and antineutrino scattering on calcium and argon are 
presented in Figs. 5 and 6, respectively, which show $d\sigma/dQ^2$ as 
functions of $Q^2$ scaled with the number of neutron/proton in the target 
(cross section per neutron/proton). Here, the results for calcium obtained in 
the RDWIA, are compared with cross sections calculated in the RFGM. The cross 
sections for ${}^{40}$Ca and ${}^{40}$Ar are almost identical. At the maximum 
the Fermi gas model results for neutrino (antineutrino) are higher than those 
obtained within the RDWIA. The discrepancy equals to 13\% (20\%) for 
$\varepsilon_{\nu}$=0.5 GeV and decreases to 7\% (12\%) for 
$\varepsilon_{\nu}$=2.5 GeV.
\begin{figure*}
  \begin{center}
    \includegraphics[height=16cm,width=16cm]{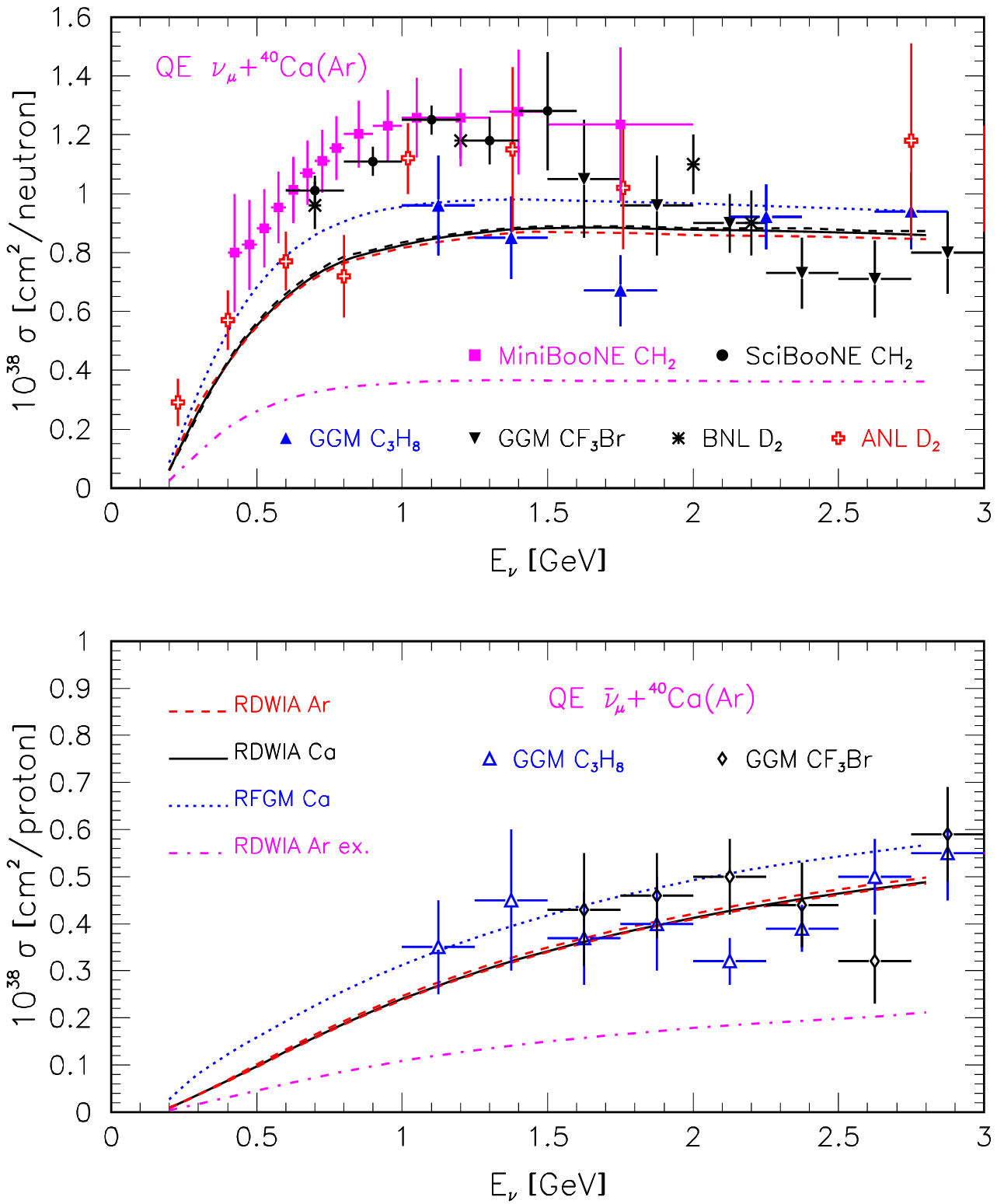}
  \end{center}
  \caption{(Color online) Total cross section  for CCQE scattering 
of muon neutrino (upper panel) and antineutrino (lower panel) on $^{40}$Ca and 
 $^{40}$Ar as a function of incoming (anti)neutrino energy. The solid and 
dashed lines are the RDWIA results for calcium and argon, respectively, while 
the dotted line is the RFGM calculation. The dashed-dotted line is the RDWIA 
result for exclusive $(\nu,\mu N)$ reactions. Data points for different targets 
are from Refs.\cite{Mann, Baker, Pohl, Brunn, MiniB,Sci1}}. 
\end{figure*}
The neutrino and antineutrino total cross sections for CCQE scattering off 
${}^{40}$Ca and ${}^{40}$Ar, calculated in the RDWIA and RFGM 
approaches with $M_A$=1.032 GeV are shown in Fig. 7 together with data from 
Refs.\cite{Mann, Baker, Pohl, Brunn, MiniB, Sci1}. Also shown are the total  
cross sections of the exclusive single-nucleon knock out $(\nu_{\mu},\mu^- p)$ 
and $(\bar{\nu}_{\mu},\mu^+ n)$ channels for (anti)neutrino scattering from 
on-shell nucleons. These reactions are source of the CCQE two-particle events 
in the final states. The cross sections are scaled with the number of 
neutron/proton in the target. 

The ratio between the neutrino cross sections calculated in the RFGM and RDWIA 
decreases with neutrino energy from about 1.26 for $\varepsilon_{\nu}$=0.3 GeV 
to $\approx$1.16 for $\varepsilon_{\nu}$=1 GeV and down to $\approx$1.09 for 
$\varepsilon_{\nu}$=2.4 GeV. For antineutrino cross sections this ratio is 
about 2.17 for $\varepsilon_{\nu}$=0.3 GeV, 1.3 for $\varepsilon_{\nu}$=1 GeV 
and 1.16 for $\varepsilon_{\nu}$=2.4 GeV. The calculated results show 
significant nuclear-model dependence for energy about 2 GeV or less.

From the experimental data shown in Fig.7 one can conclude that the CCQE total 
cross sections measured in different experiments can vary by 20-40\%. The data
 have large systematic uncertainties due to the pour knowledge of the 
background contamination in selected events and/or the incoming neutrino 
flux. Obtaining a reliable estimate of the neutrino flux is notoriously 
difficult and remains a challenge.

Selection techniques that rely on the identification of a single final state 
proton (two track CCQE events) can improve significantly the purity of the QE 
sample. Moreover, a simultaneous measurement of both two track and single muon 
track events~\cite{NOMAD} allow to constrain the systematics associated with 
the FSI and the neutrino flux. The ratio of the exclusive $(\nu_{\mu},\mu^- p)$ 
reaction total cross section to the CCQE total cross section is an attractive 
quantity because it is supposed to be rather insensitive to the neutrino flux 
uncertainty and can be especially susceptible to the FSI effects.

We calculated the $R^{ex}=\sigma^{ex}_{tot}/\sigma_{tot}$ ratio, where 
$\sigma^{ex}_{tot}$ is the total cross section of the $(\nu_{\mu},\mu N)$ 
reaction for the (anti)neutrino scattering on shell-nucleons on the 
carbon~\cite{BAV4}, oxygen~\cite{BAV2} and argon. The results calculated in 
the RDWIA are shown in Fig.~8 as functions of (anti)neutrino energy. This 
figure clearly shows that the ratio reduces with the mass number of the target.
 The function $R^{ex}(\varepsilon_{\nu})$ has a maximum in the range 
$\varepsilon_{\nu}$=0.3-0.4 GeV and decreases slowly with neutrino energy. For 
carbon (oxygen) (argon) the ratio is 0.59 (0.55) (0.48) at the maximum and 
$\approx$0.54 (0.46) (0.43) at $\varepsilon_{\nu}$=2.8 GeV. So, due to the FSI 
the contribution of the exclusive channels reduces slowly with the neutrino 
energy and mass-number of the target.
\begin{figure*}
  \begin{center}
    \includegraphics[height=16cm,width=16cm]{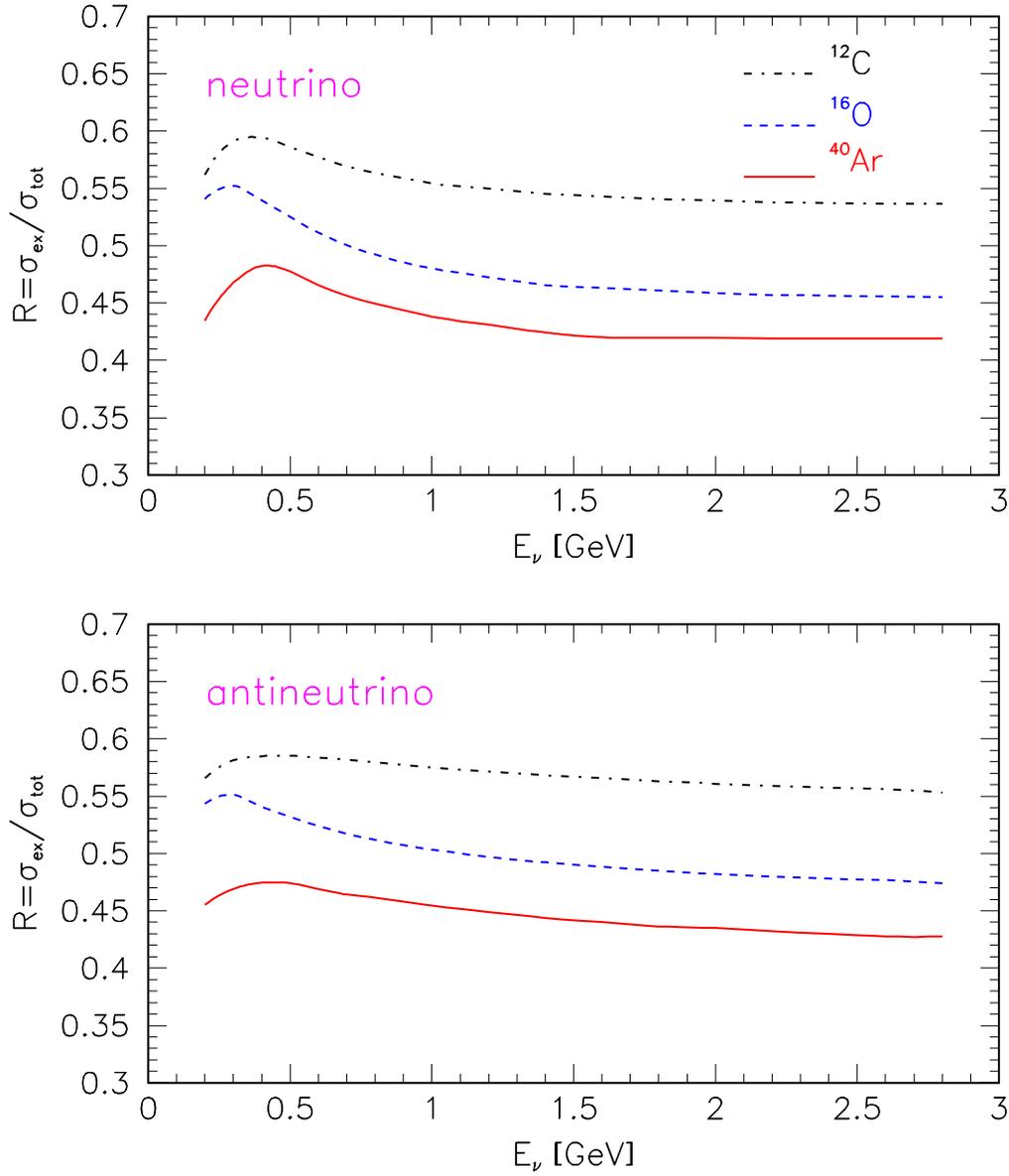}
  \end{center}
  \caption{(Color online) Ratio $R^{ex}$ of the $(\nu_{\mu},\mu N)$ reaction 
total cross section to the CCQE total cross section for muon neutrino 
(upper panel) and antineutrino (lower panel) scattering on ${}^{12}$C 
(dashed-dotted line), ${}^{16}$O (dashed line), and ${}^{40}$Ar (solid line) vs
 incoming (anti)neutrino energy. The ratio was calculated in the RDWIA 
approach.} 
\end{figure*}

Most long base-line neutrino oscillation experiments try to reduce systematic 
errors in the extraction of the oscillation parameters by using near and far 
detectors. Because these detectors are not necessarily of the same 
target material we estimated the difference between the total cross sections 
per (proton)neutron for the (anti)neutrino CCQE scattering off ${}^{12}$C, 
${}^{16}$O, and ${}^{40}$Ar. The ratio
$R(\varepsilon_{\nu})=(\sigma^{Ar}_{tot})_{nucl}/(\sigma^{C}_{tot})_{nucl}$ 
($Ar/C$ ratio) was calculated, where the cross sections 
$(\sigma^i_{tot})_{nucl}$ are scaled with the number of neutron/proton in the 
target. The results obtained in the RFGM and RDWIA are shown in 
Fig.9. The ratio 
$R(\varepsilon_{\nu}=(\sigma^{O}_{tot})_{nucl}/(\sigma^{C}_{tot})_{nucl}$ 
($O/C$ ratio) calculated in Ref.~\cite{BAV4} is also shown for comparison.

The Fermi gas model predicts almost identical values of $\sigma^O_{tot}$ and 
$\sigma^C_{tot}$. For neutrino (antineutrino) scattering the ratio $Ar/C$
increases from 0.89(0.84) at $\varepsilon_{\nu}$=0.3 GeV up to 0.99(0.98) at 
$\varepsilon_{\nu}$=2.8 GeV. In the RDWIA approach the ratios $O/C$ and $Ar/C$ 
are lower than those calculated in the RFGM. For the neutrino(antineutrino) 
scattering $O/C$ is 0.88(0.85) at $\varepsilon_{\nu}$=0.5 GeV and increases 
slowly with energy up to 0.93(0.92) at $\varepsilon_{\nu}$=2.8 GeV. On the other
 hand $Ar/C$ ratio of 0.95(0.75) at $\varepsilon_{\nu}$=0.5 GeV decreases with 
energy up to 0.88 for the neutrino interaction and increases up to 0.88 for 
the antineutrino scattering at $\varepsilon_{\nu}$=2.8 GeV.
\begin{figure*}
  \begin{center}
    \includegraphics[height=16cm,width=16cm]{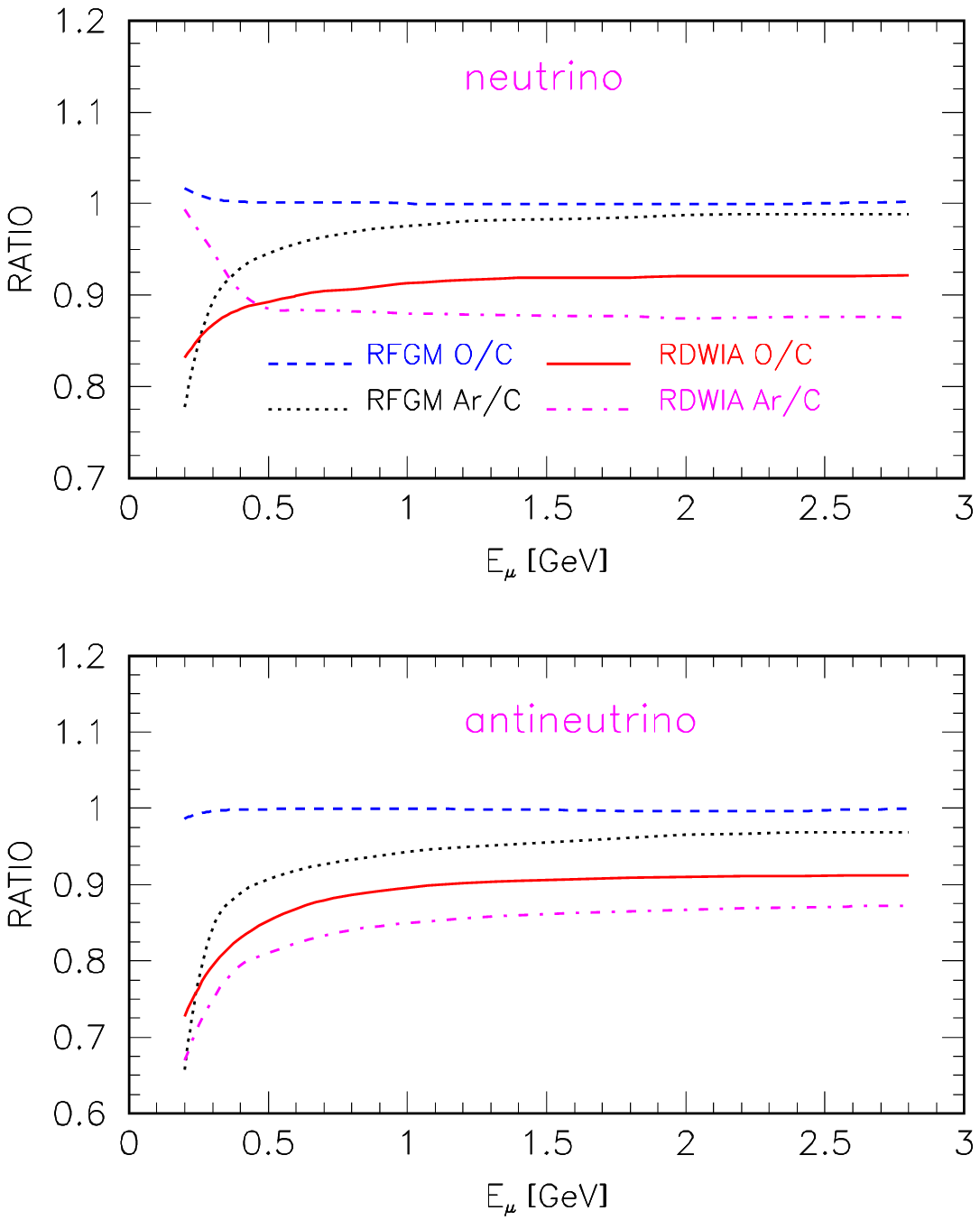}
  \end{center}
  \caption{(Color online) Ratio of the total cross sections per neutron/proton 
$R=O/C$ (solid and dashed lines)~\cite{BAV4} and $R=Ar/C$ (dashed-dotted and 
dotted lines) for CCQE scattering of muon neutrino (upper panel) and 
antineutrino (lower panel) as a function of incoming (anti)neutrino energy. As 
shown in the key, the cross sections were calculated with the RDWIA and RFGM 
approaches.} 
\end{figure*}
\begin{figure*}
  \begin{center}
    \includegraphics[height=16cm,width=16cm]{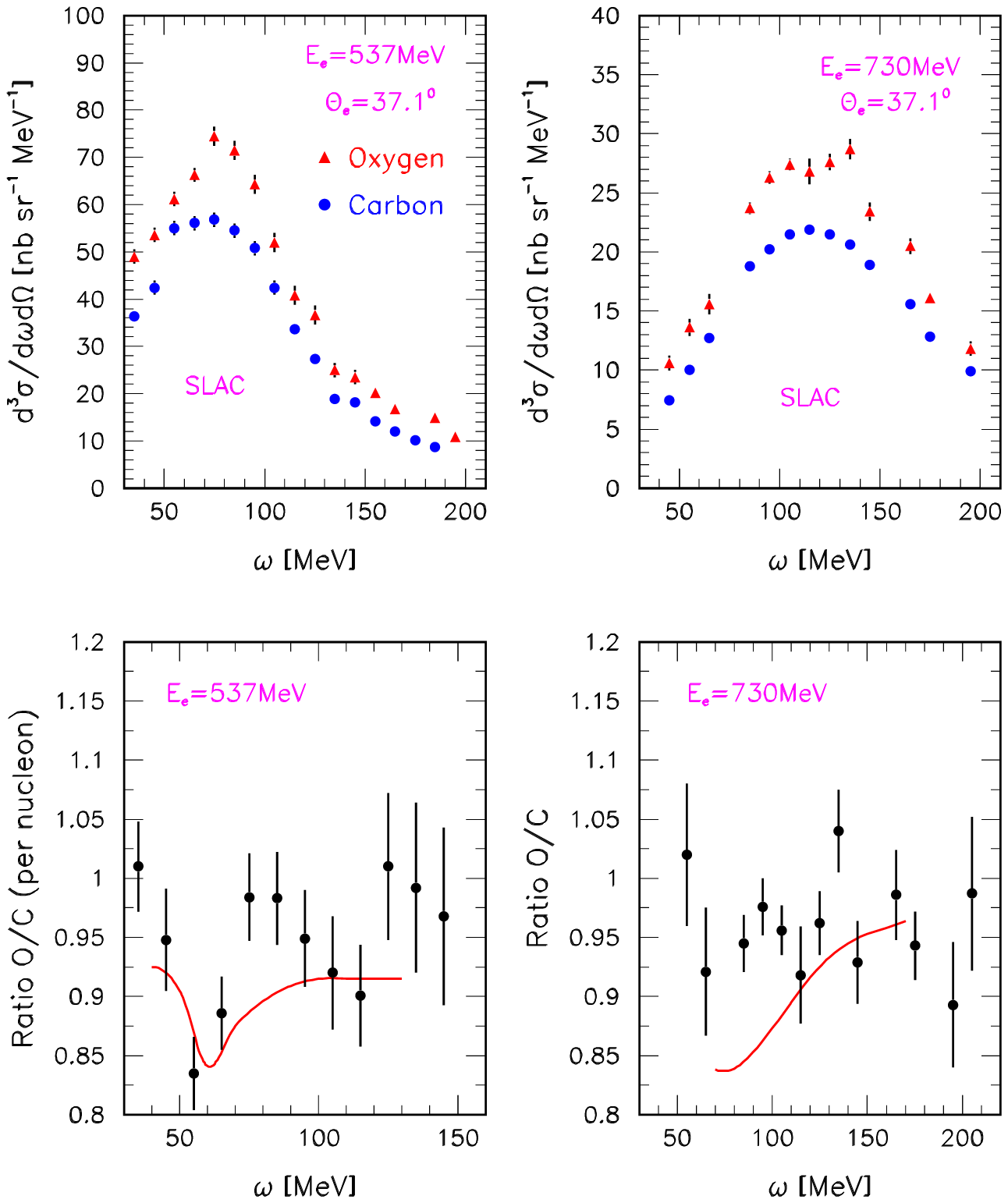}
  \end{center}
  \caption{(Color online) Inclusive cross sections (upper panel) per nucleon 
and ratio $O/C$ (lower panel) vs energy transfer $\omega$ for electron 
scattering on ${}^{12}$C and ${}^{16}$O. Data for carbon (filled circles) 
and oxygen (filled triangles) are from Ref.~\cite{OCon} for electron beam 
energies $\varepsilon_e$=537, 730 MeV and scattering angle $\theta_e$=37.1
$^{\circ}$. The solid line is the RDWIA calculation.} 
\end{figure*}
\begin{figure*}
  \begin{center}
    \includegraphics[height=16cm,width=16cm]{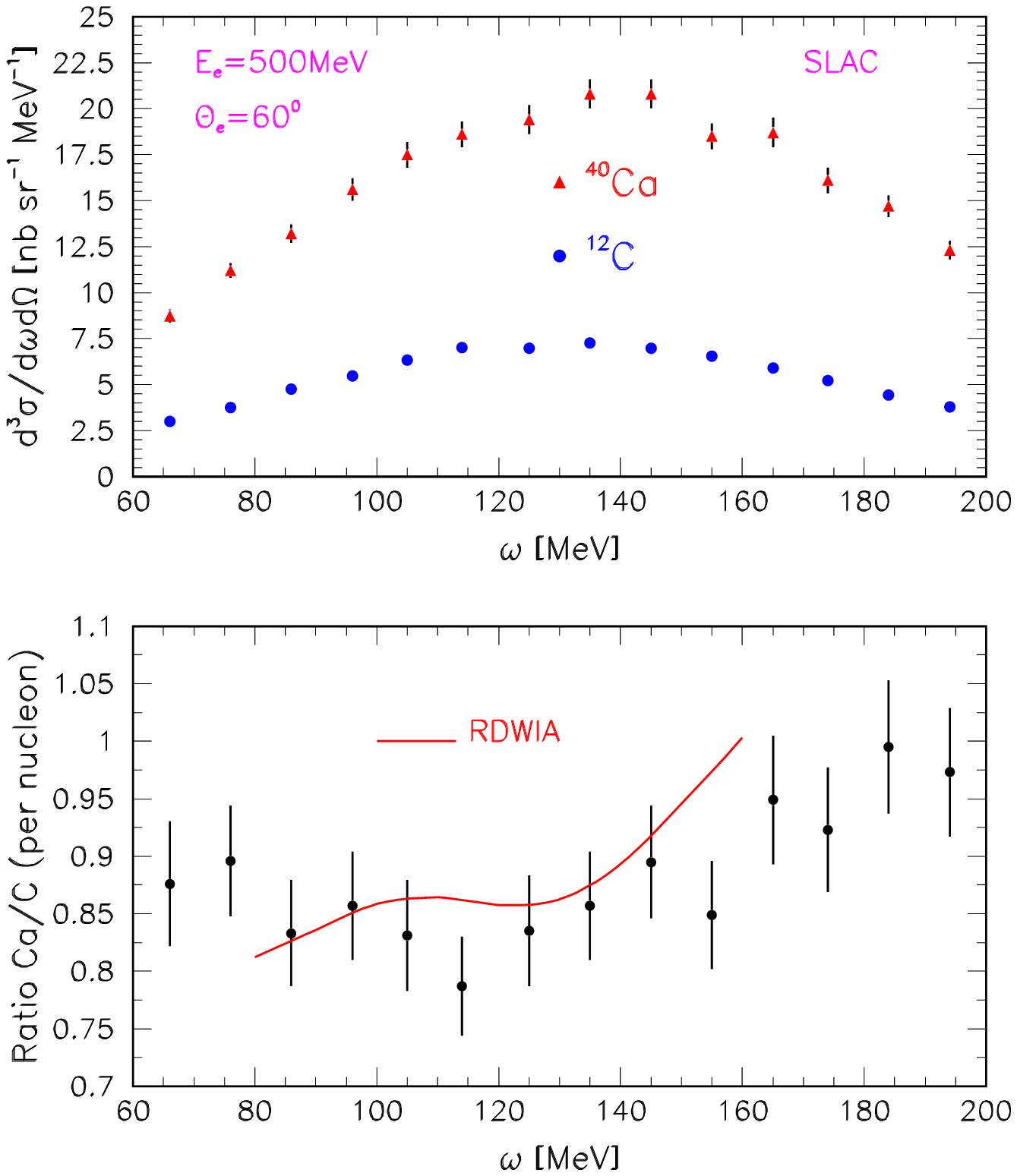}
  \end{center}
  \caption{(Color online) Inclusive cross sections (upper panel) per nucleon 
and ratio $Ca/C$ (lower panel) vs. energy transfer $\omega$ for electron 
scattering on ${}^{12}$C and ${}^{40}$Ca. Data for carbon (filled circles) 
and calcium (filled triangles) are from Ref.~\cite{Whit} for electron beam 
energies $\varepsilon_e$=500 MeV and scattering angle $\theta_e$=60$^{\circ}$. 
The solid line is the RDWIA calculation.} 
\end{figure*}

In the SLAC experiments the inclusive cross sections 
$d\sigma/d\varepsilon d\Omega$ for electron scattering on ${}^{12}$C and 
${}^{16}$O~\cite{OCon} as well as on ${}^{12}$C and ${}^{16}$O~\cite{Whit} were 
measured in the same kinematical conditions. Using these data we calculated the 
$(O/C)_{el}=(d\sigma^O/d\varepsilon d\Omega)_{nucl}/(d\sigma^C/
d\varepsilon d\Omega)_{nucl}$ and 
$(Ca/C)_{el}=(d\sigma^{Ca}/d\varepsilon d\Omega)_{nucl}/(d\sigma^C/
d\varepsilon d\Omega)_{nucl}$ ratios, where the differential cross sections 
($d\sigma^i/d\varepsilon d\Omega)_{nucl}$ are scaled with the number of nucleons
 in the targets. Figures 10 and 11 show the measured ratios as functions of 
energy transfer as compared to the RDWIA calculations in the QE peak region.

There is an agreement between the RDWIA results and the $(O/C)_{el}$ data 
within the error of the experiments, whereas the uncertainties of the measured 
$(O/C)_{el}$ ratios of 5-10\% are the same order as the predicted effects. On 
the other hand the calculated $(Ca/C)_{el}$ ratio agree well with data where the
 observed effect of 15\% in the QE peak region is higher than experimental 
errors. Thus the RDWIA model predicts that due to nuclear effects the CCQE 
differential and total cross sections per neutron/proton reduces with the mass
 number of the targets.

To investigate why the (anti)neutrino CCQE total cross sections per 
neutron/proton for ${}^{12}$C, ${}^{16}$O, and ${}^{40}$Ar are dissimilar we 
estimated the nuclear structure, short-range $NN$ correlation, and FSI effects.

The nuclear structure effects are different in
${}^{16}$O(${}^{40}$Ar) and ${}^{12}$C cross sections due to different nucleon 
binding energies and momentum distributions for all bound nucleon states in 
these nuclei. To estimate these effects we calculated the ratios 
$R_{str}(\varepsilon_{\nu})=O/C(Ar/C)$ in the PWIA approach without the $NN$ pair 
contributions, assuming that the occupancy of all nuclear shells $S_{\alpha}$=1
in carbon, oxygen and argon. The nuclear structure effect was estimated 
as $\Delta_{str}$=$1-R_{str}$. Then we calculated in the same approach the 
$R_{NN}(\varepsilon_{\nu})$ ratios in the presence of the $NN$ correlations in 
the ground states of nuclei. Note, in our calculations the averaged 
occupancies of the IPSM orbitals of ${}^{12}$C, ${}^{16}$O, and ${}^{40}$Ar are: 
$S_C$=0.89, $S_O$=0.75, and $S_{Ar}$=0.87. The $NN$ correlation effect, i.e. 
the difference between the cross sections as a consequence of different $NN$ 
pair contributions was calculated as $\Delta_{NN}$=$R_{str} - R_{NN}$. The FSI 
effects are different between the cross sections due to interaction of 
the outgoing nucleons with the different residual nuclei. For example, 
$p + {}^{11}$C for the neutrino or $n + {}^{11}$B for the antineutrino 
scattering off the ${}^{12}$C and $p + {}^{15}$O for the neutrino and 
$n + {}^{15}$N for the antineutrino scattering off ${}^{16}$O. These effects 
were estimated as the difference $\Delta_{FSI}$=$R_{NN} -R $, where the ratio 
$R$ is calculated in the RDWIA approach is shown in Fig. 9 and 10. The total 
nuclear effect is sum 
$\Delta_{tot}$=$\Delta_{str} + \Delta_{NN} + \Delta_{FSI}=1 - R$. 
\begin{figure*}
  \begin{center}
    \includegraphics[height=16cm,width=16cm]{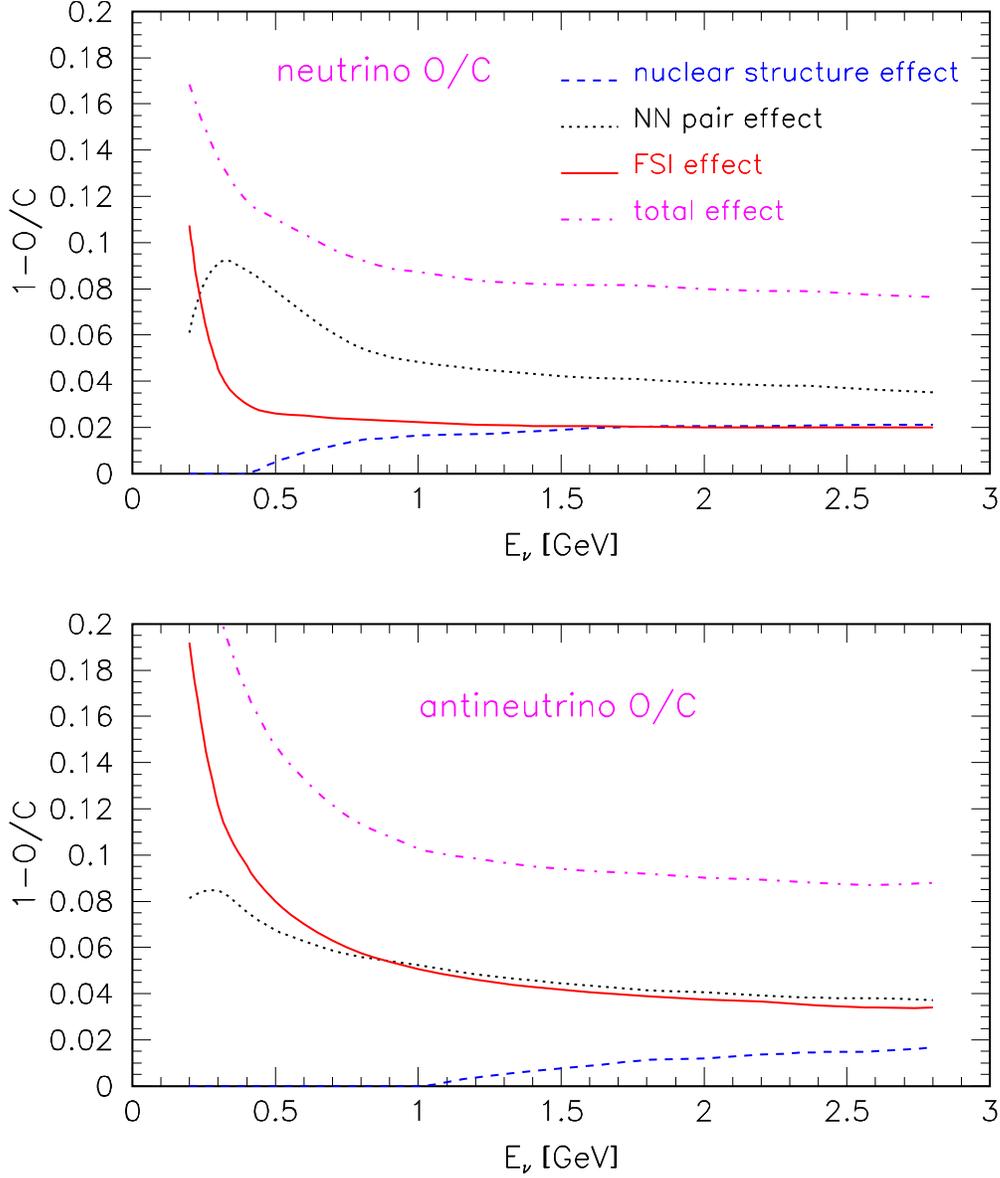}
  \end{center}
  \caption{(Color online) Nuclear structure (dashed line), $NN$ correlation 
(dotted line), FSI (solid line), and total nuclear effects (dashed-dotted) 
(see text) for CCQE scattering of muon neutrino (upper panel) and antineutrino
 (lower panel) on ${}^{12}$C and ${}^{16}$O vs incoming (anti)neutrino energy.}
\end{figure*}
\begin{figure*}
  \begin{center}
    \includegraphics[height=16cm,width=16cm]{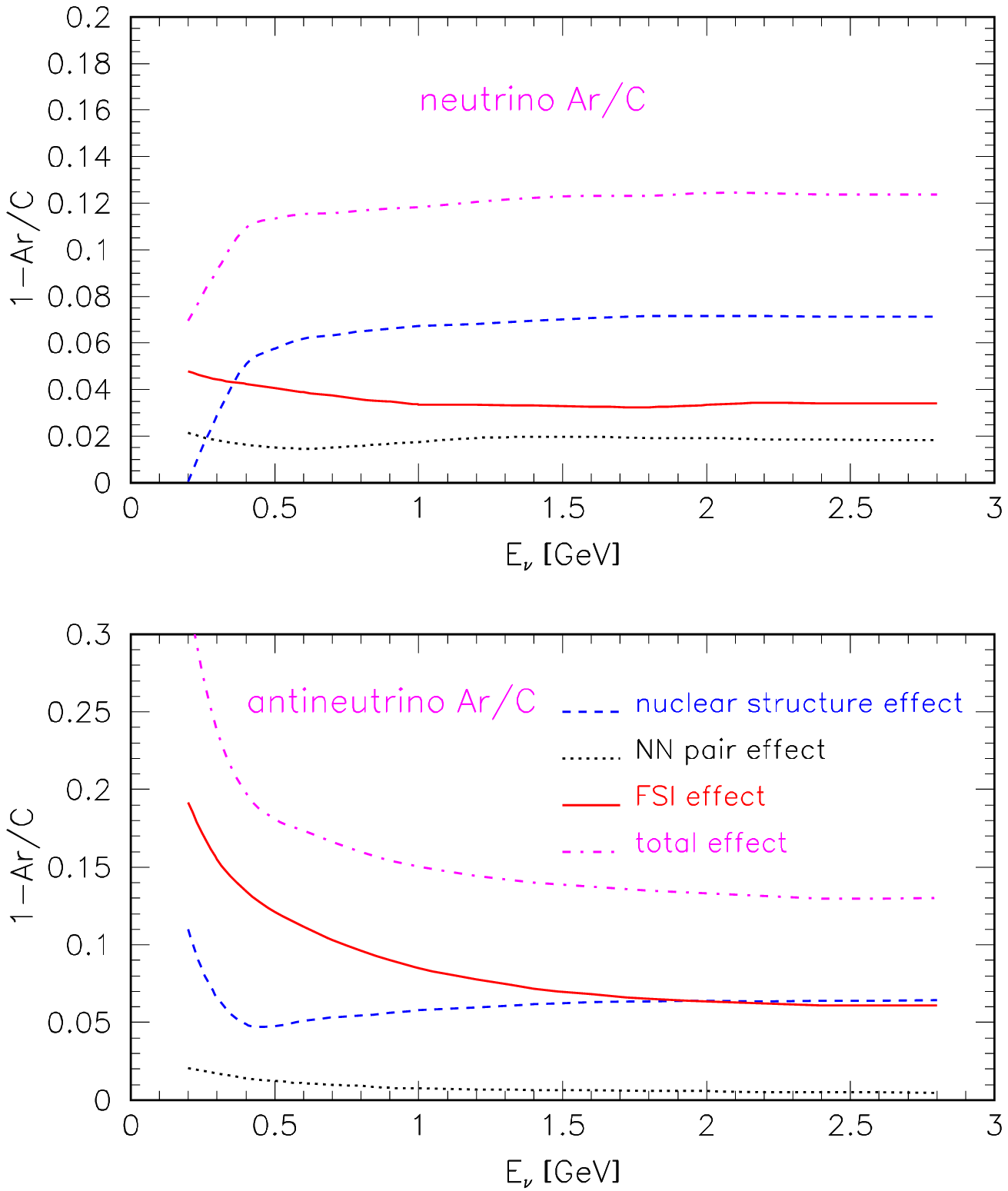}
  \end{center}
  \caption{(Color online) Same as Fig.12 but for (anti)neutrino scattering on 
${}^{12}$C and ${}^{40}$Ar.} 
\end{figure*}

The result of the cross section comparison for the (anti)neutrino scattering on 
carbon and oxygen as well as on carbon and argon is displayed in 
Figs.~12 and 13, correspondingly. In these figures the differences 
$\Delta_{str}, \Delta_{NN}, \Delta_{FSI}$ and $\Delta_{tot}$ are shown as 
functions of neutrino energy. Comparison of the cross sections for 
${}^{12}$C and ${}^{16}$O shows that the nuclear structure effects for the 
neutrino and antineutrino interactions are small and similar. The function 
$\Delta_{str}$ increases slowly with energy up to $\approx$2\% at 
$\varepsilon_{\nu}$=2.8 GeV. The $NN$ correlation effects decrease with increased
(anti)neutrino energy. For the neutrino(antineutrino) scattering the function 
$\Delta_{NN}(\varepsilon_{\nu})$ reduces from $\approx$ 8\%(7\%) at 
$\varepsilon_{\nu}$=0.5 GeV up to $\approx$4\%(4\%) at 
$\varepsilon_{\nu}$=2.8 GeV. The FSI effects are higher for the antineutrino 
scattering than those for the neutrino interaction. For the neutrino 
(antineutrino) scattering the function $\Delta_{FSI}(\varepsilon_{\nu})$ 
decreases from $\approx$3\%(9\%) at $\varepsilon_{\nu}$=0.5 GeV down to 
$\approx$2\%(3.5\%) at $\varepsilon_{\nu}$=2.8 GeV and the total nuclear effect 
$\Delta_{tot}(\varepsilon_{\nu})$=0.11(0.17) at $\varepsilon_{\nu}$=0.5 GeV and 
reduces down to 0.08(0.09) at $\varepsilon_{\nu}$=2.8 GeV.

To conclude, the main sources of the distinction between the carbon and oxygen 
cross sections are different $NN$ pair contributions in the ${}^{12}$C (11\%) 
and ${}^{16}$O (25\%) ground states. For the antineutrino cross sections the 
FSI effects are the same order as $NN$ correlation effects at 
$\varepsilon_{\nu}\ge$ 1 GeV. The precise measurement and accurate calculation 
of the $NN$ correlation contributions are important for a reliable estimation 
of the difference between the ${}^{12}$C and ${}^{16}$O cross sections.    

Comparison of the cross sections for ${}^{12}$C and ${}^{40}$Ar shows that 
the nuclear structure effects at $\varepsilon_{\nu} \ge$0.5 GeV are identical 
for neutrino and antineutrino interactions. The function $\Delta_{str}$=0.06 at 
$\varepsilon_{\nu}$=0.5 GeV and increases slowly with energy up to $\approx$0.07
 at $\varepsilon_{\nu}$=2.8 GeV. The $NN$ correlation effects of 1-2\% are 
small. The FSI effects are higher for the antineutrino scattering. For the 
neutrino (antineutrino) interaction $\Delta_{FSI}$ is about 0.045(0.13) for 
$\varepsilon_{\nu}$=0.5 GeV and reduces with energy down to 0.035(0.06) for     
$\varepsilon_{\nu}$=2.8 GeV and the total nuclear effect $\Delta_{tot}$=0.12
(0.18) at $\varepsilon_{\nu}$=0.5 GeV and $\approx$0.13(0.13) at 
$\varepsilon_{\nu}$=2.8 GeV. Thus the nuclear structure and FSI effects give the
 dominant contribution to the difference between the ${}^{12}$C and ${}^{40}$Ar
 total cross sections per neutron/proton.    

\section{Conclusions}

In this work, we study electron and CCQE (anti)neutrino
scattering on calcium and argon targets in different approaches (PWIA, 
RDWIA, RFGM). The RDWIA model were widely and successfully applied to the 
analysis of the available electron scattering data over a wide range of nuclei.

First, the reduced cross sections for electron and (anti)neutrino scattering 
off ${}^{40}$Ca calculated in the RDWIA were tested against ${}^{40}$Ca$(e,e'p)$
 data. We found that the result for (anti)neutrino scattering are similar to 
those for electron scattering (apart from the small differences due to the 
Coulomb correction) and the latter are in good agreement with the electron 
data. Also it was shown that the reduced cross sections for the removal of the 
proton from the shells of ${}^{40}$Ca and ${}^{40}$Ar are very similar.

The inclusive cross sections, calculated in the RDWIA model, which has been 
modified with phenomenological spectroscopic factors and nucleon high-momentum 
components in the target were tested against ${}^{40}$Ca$(e,e')$ data. The 
results generally agree within about 14\%. On the other hand, in the QE peak 
region the RFGM overestimates the value of the inclusive cross sections at low 
momentum transfer and the discrepancy with data reduces as this momentum 
increases. Also, it was shown that the measured and calculated in the RDWIA 
inclusive cross sections per nucleon for electron scattering off ${}^{12}$C, 
${}^{16}$O, and ${}^{40}$Ca decreases with the mass-number of the target in the 
QE peak region.

The CCQE total cross sections for (anti)neutrino scattering on calcium and 
argon predicted by the RFGM are higher than those obtained in 
the RDWIA and the difference decreases with (anti)neutrino energy. The relative 
contribution of the $(\nu_{\mu}, \mu N)$ channels to the CCQE total cross section
 is lower for heavier nuclei due to the FSI and $NN$ pair effects. For 
(anti)neutrino scattering on carbon, oxygen, and argon we 
compared the total cross sections (scaled with the number of neutrons/protons in
 the target) and found that the cross sections calculated within the RDWIA for 
${}^{16}$O and  ${}^{40}$Ar are lower than those calculated for ${}^{12}$C. In 
the RFGM the cross sections for carbon are practically equal to those for 
oxygen and they are higher by about 2-5\% than those for argon at 
$\varepsilon \ge$0.5 GeV.  

We also studied different sources of the distinction between the 
${}^{12}$C, ${}^{16}$O, and ${}^{40}$Ar cross sections per neutron/proton. We have
found that the difference between the ${}^{12}$C and ${}^{16}$O cross sections 
is mainly due to different $NN$ correlation contributions in the ground states
 of the nuclei and FSI effects. The main sources of the difference between the
 ${}^{12}$C and ${}^{40}$Ar cross sections are the nuclear structure and FSI 
effects.    

Thus the RDWIA approach predicts that the CCQE differential and total cross 
sections per neutron/proton reduces slowly with the mass-number of the target 
due to the nuclear effects. Although the model and theoretical ingredients 
adopted in the calculations contain approximations, our results can serve as
 a useful reference for long base-line neutrino oscillation experiments

\section*{Acknowledgments}

We thank L. Lapik\'{a}s for providing the NIKHEF data tables. We especially 
thank J. Morfin, B. Ziemer and D. Perevalov for fruitful discussions and a 
critical reading of the manuscript. The support of the Russian Academy of 
Science is gratefully acknowledged. 
%


\end{document}